\documentclass[onecolumn]{elsart3p}
\newcommand{\rtinv}{{\rm Re}\ t^{-1} }
\newcommand{\ui}{{\rm i}\ }
\newcommand{\comp}{{\rm C}\hspace{-1ex}\rule{0.1mm}{1.5ex}\hspace{1ex}}
\begin{document}

\begin{frontmatter}

\title{Meson Resonances at large $N_C$: Complex Poles vs
  Breit-Wigner Masses}~\thanks{
Research Supported by DGI and FEDER funds, under contracts
FIS2008-01143/FIS and the Spanish Consolider-Ingenio 2010 Programme
CPAN (CSD2007-00042), by Junta de Andaluc\'\i a under contract
FQM0225. It is part of the European Community-Research Infrastructure
Integrating Activity ``Study of Strongly Interacting Matter'' (acronym
HadronPhysics2, Grant Agreement n. 227431) and of the EU Human
Resources and Mobility Activity ``FLAVIAnet'' (contract number
MRTN--CT--2006--035482) , under the Seventh Framework Programme of EU.}

\author{J.  Nieves}
\address{ Instituto de F\'\i sica Corpuscular (IFIC), Centro Mixto
  CSIC-Universidad de Valencia, Institutos de Investigaci\'on de
  Paterna, Aptd. 22085, E-46071 Valencia, Spain} 

\author{E. Ruiz  Arriola}
\address{Departamento de
  F\'{\i}sica At\'omica, Molecular y Nuclear, Universidad de Granada,
  E-18071 Granada, Spain.}


\date{\today}

\begin{abstract}
The rigorous quantum mechanical definition of a resonance requires
determining the pole position in the second Riemann sheet of the
analytically continued partial wave scattering amplitude in the
complex Mandelstam $s$ variable plane. For meson resonances we
investigate the alternative Breit-Wigner (BW) definition within the
large $N_C$ expansion.  By assuming that the pole position is ${\cal
O} (N_C^{0})$ and exploiting unitarity, we show that the BW
determination of the resonance mass differs from the pole position by
${\cal O} (N_C^{-2})$ terms, which can be extracted from $\pi\pi$
scattering data.  For the case of the $f_0(600)$ pole, the BW scalar mass is
predicted to occur at $\sim 700\ {\rm MeV}$ while the true value is
located at $\sim 800\ {\rm MeV}$.
\end{abstract}


\end{frontmatter}
Meson resonances are key building blocks in intermediate energy
hadronic physics (for a review see e.g. Ref.~\cite{Badalian:1981xj}
and references therein). Most often they contribute as virtual
intermediate states to physical processes. This poses the question on
the suitable interpolating field since, when the resonance goes
off-shell, a definition of the background becomes necessary and its
non-elementary nature becomes evident (see
e.g.~\cite{Ecker:1988te,Kampf:2006yf}).  The large $N_C$ expansion of
QCD~\cite{'tHooft:1973jz,Witten:1979kh} provides a handle on this
problem since meson resonances with a $q\bar q$ component, dominant or
sub-dominant for $N_C=3$, become stable particles; their mass becomes
a fixed number $m_R \sim N_C^0$ and their width is suppressed as
$\Gamma_R \sim 1/N_C$ for a sufficient large number of colors. This 
justifies the usage of a tree level Lagrangian in terms of
canonically quantized fields~(see e.g. \cite{Pich:2002xy} and
references therein); resonance widths appear naturally as decay rates
of the classical stable particles or equivalently as a quantum
self-energy correction to the resonance propagator. Depending on the
numerical value of the mass, being above or below threshold, physical
resonances turn into Feschbach resonances or bound states respectively
(see e.g.~\cite{Jaffe:2007id}).

In general~~\cite{'tHooft:1973jz,Witten:1979kh} one expects a series
expansion of the complex pole position $s_R = m_R^2- \ui \Gamma_R m_R$,
of the $S-$matrix in the Second Riemann Sheet (SRS),
\begin{eqnarray}
s_R = s_R^{(0)} + \lambda s_R^{(1)} + \lambda^2 s_R^{(2)} + \dots
\label{eq:sRII}
\end{eqnarray} 
where $\lambda=3/N_C$.  The purpose of this note is to show that using
the standard and well-known Breit-Wigner (BW) definition with a
similar expansion
\begin{eqnarray}
s_{\rm BW} = s_{\rm BW}^{(0)} + \lambda s_{\rm BW}^{(1)} + 
\lambda^2 s_{\rm BW}^{(2)} + \dots  
\end{eqnarray} 
one has that 
\begin{eqnarray}
s_{\rm BW} -{\rm Re} (s_{\rm R} )= {\cal O} (N_C^{-2}) \, , 
\end{eqnarray} 
anticipating an improved convergence and also suggesting a model
independent way of assessing the accuracy of the large $N_C$ expansion.

Large $N_C$ scaling away from the physical $N_C=3$ value, but
relatively close to it, has been applied to chiral unitarized $\pi\pi$
amplitudes in Refs.~\cite{Pelaez:2003dy,Pelaez:2006nj} as a method to
learn on the nature of meson resonances and on the induced $N_C$ dependence
of the corresponding pole masses and widths. We have recently
shown~\cite{Nieves:2009ez} that in some cases, as for instance that of
the $f_0(600)$ resonance, there is a lack of predictive power on the
true $N_C$ behaviour of the pole in the $N_C\to \infty$ limit, which
could only be fixed by fine-tuning the parameters to unrealistically
precise values. Two loop unitarized calculations are, in addition,
beset by large uncertainties~\cite{Nieves:2001de}. Though meaningful
consequences can be drawn by studying the behaviour of the resonance
in the vicinity of $N_C=3$, we do not share the view~\cite{Sun:2005uk}
that one can reliably follow the $N_C$ trajectory far from the real
world ($N_C=3$), extrapolating from calculations which are
phenomenologically successful at $N_C=3$, mainly because large
uncertainties are built in. In addition to spurious $1/N_C$
corrections, the amplitude may not contain all possible leading $N_C$
terms which are relevant at the resonance energies when $N_C$ grows.
We believe instead that more robust results might be achieved by
examining observables which are {\it parametrically} suppressed by
$1/N_C^2$, rather than just by $1/N_C$ corrections, but keeping always
$N_C=3$. This is in fact the way how the large $N_C$ expansion has
traditionally proven to be most
powerful~\cite{Manohar:1998xv,Jenkins:1998wy}.

Let us consider for definiteness elastic $\pi\pi$ scattering in a given
isospin--angular momentum sector denoted as $(T,J)$, and let us also
neglect coupled channel effects. The $S-$matrix is defined as
\begin{eqnarray} 
S_{\rm TJ} (s) = e^{2 {\rm i} \delta_{TJ}(s)}= 1 - 2
\ui \rho(s) t_{\rm TJ} (s) \, ,\, s \ge 4 m^2
\label{eq:s-matrix}
\end{eqnarray}
with $s$ the total $\pi\pi$ center of mass energy, $\delta_{TJ}(s)$
the phase shift, $t_{\rm TJ} (s)$ the scattering amplitude, $m$ the
pion mass and
\begin{eqnarray} 
\rho (s) = 
\frac{1}{16\pi}\sqrt{1-\frac{4 m^2}s} \, ,\, s \ge 4 m^2
\end{eqnarray} 
the phase space in our particular normalization.  For simplicity we
will drop the partial wave channel $(T,J)$ indices in what follows.
Using Eq.~(\ref{eq:s-matrix}) we deduce
\begin{eqnarray}
\tan\left[\delta(s)-\frac{\pi}{2}\right] = \frac{\rtinv
  (s)}{\rho(s)}\, ,\, s \ge 4 m^2 
\, .
\end{eqnarray}
Let us write 
the large $N_C$ expansion of the partial wave amplitude
\begin{eqnarray}
t(s) = \lambda t_1 (s)  + \lambda^2 t_2 (s) + \lambda^3 t_3(s) + \dots  
\end{eqnarray} 
where the $t_n(s)$ are taken as $N_C$
independent. From two-particle unitarity, which  we write in the inverse
amplitude form as
\begin{eqnarray}
t(s)^{-1} = {\rm Re} \, t(s)^{-1} + {\rm i}\ \rho(s) \, ,  
\end{eqnarray} 
we get the constraints 
\begin{eqnarray}
{\rm Im} \, t_1(s) &=& 0 \, , \\    
{\rm Im} \, t_2(s) &=& -\rho(s) \, t_1^2(s) \, , \\
{\rm Im} \, t_3(s)
 &=& -2\rho(s) t_1(s) \, {\rm Re} t_2 (s)  \, ,   
\end{eqnarray} 
and so on. Note that the leading $N_C$ amplitude is real in the
elastic scattering region, as expected from a tree level $\pi\pi$
amplitude~\cite{'tHooft:1973jz,Witten:1979kh}. Of course, this does
not preclude the appearance of the left cut discontinuity which occurs
due to particle exchange in the $t$ and $u$ channels. Clearly any
pole, $s_0$, occurring for the leading $N_C$ and {\it real } amplitude
will be either real or occurs in complex conjugated pairs. The latter
is excluded as this would violate causality. If $s_0 < 4 m^2 $ it
corresponds to a bound state while for $s_0> 4 m^2$ it can be
associated to a Feschbach resonance.


To analytically continue the scattering amplitude to the complex
Mandelstam $s-$plane, we remind that above threshold, elastic
unitarity fixes the imaginary part of the inverse of the $t-$matrix,
which is then determined as the boundary value in the upper lip of the
unitarity cut,
\begin{eqnarray}
t^{-1}(s+\ui \epsilon) &=& \rtinv (s) + \ui {\cal R}(s+\ui \epsilon),
\nonumber \\ {\cal R}(s+\ui \epsilon) &\equiv& \rho(s) \ge 0, \quad s
\ge
4m^2 \, . \label{eq:tinvI}
\end{eqnarray}
%
%
Resonances manifest as poles in the fourth quadrant of the SRS of the
$t-$matrix. The $t-$matrix in the First Riemann Sheet (FRS), $ t_{\rm
  I}$, is defined in the complex plane by means of an analytical
continuation of its boundary value in Eq.~(\ref{eq:tinvI}) at the
upper lip of the unitarity cut.  The $t-$matrix in the SRS ($t_{\rm
  II}$) is related to $t_{\rm I}$, thanks to ${\cal R}(s+\ui
\epsilon)= - {\cal R}(s-\ui \epsilon)$, by~\cite{Nieves:2001wt}
\begin{eqnarray}
t^{-1}_{II}(z) = t^{-1}_I(z) - 2\ui {\cal R}(z), \quad z \in \comp \, , 
\end{eqnarray}
which implements continuity through the unitarity right cut, and the
requirement that there are only two Riemann sheets associated to this
cut,
\begin{eqnarray}
t^{-1}_{II}(s \mp \ui\epsilon) = t^{-1}_I(s \pm \ui\epsilon), \quad s
\ge 4m^2 \, . 
\end{eqnarray}
Let $s_R=m^2_R-\ui m_R \Gamma_R$, the position of the pole associated
to the resonance $R$. By definition $s_R$, it is solution of the
equation $t^{-1}_{II} (s_R)=0$, which can be expressed as 
\begin{eqnarray}
\rtinv_I (s_R) = -\ui {\cal R}(s_R^*) \, , 
\label{eq:resonance}
\end{eqnarray}
where we have used that ${\cal R}(s_R)=-{\cal R}(s_R^*)$\footnote{We
  are being abusive regarding notation. Here $\rtinv_I (z)$ is an
  analytical function which has not right cut and it does correspond to
  the real part of a function {\it only} when $z=s+\ui \epsilon$.}.


In the large $N_C$ limit, $\rtinv_I$ and ${\cal R}$ scale as ${\cal O}
(N_C)$ and ${\cal O} (N_C^0)$, respectively, and thus one easily finds
that $m_R$ and $\Gamma_R$ do scale as ${\cal O} (N_C^0)$ and ${\cal O}
(N_C^{-1})$, respectively as we now show. Indeed, the resonance pole
position, $s_R$, satisfies Eq.~(\ref{eq:resonance}), and propose $N_C$
expansions of the type (for simplicity, we will drop out the sub-index
$I$, associated to the FRS)
\begin{eqnarray}
s_R &=& x_R + \frac{y_R}{N_C} + {\cal O} (N_C^{-2}) \, , \\ 
\rtinv &=& \left( \rtinv \right)_{(1)} + \left( \rtinv
\right)_{(0)} +
       {\cal O} (N_C^{-1})\, , 
\end{eqnarray}
where we have used that any pole generated by the re-summation of
diagrams must necessarily scale as ${\cal O} (N_C^0)$ for a
sufficiently large number of colors and that
$\rtinv$ scales as ${\cal O} (N_C)$ (we use an obvious notation in
the $N_C$ expansion of $\rtinv$, where $\left( \rtinv \right)_{(j)}$ scales
as ${\cal O} (N_C^j)$).  The large $N_C$ expansion of
Eq.~(\ref{eq:resonance}) reads
\begin{eqnarray}
&& \underbrace{\left( \rtinv \right)_{(1)}(x_R)}_{{\cal O} (N_C)} +
  \underbrace{\frac{y_R}{N_C} \left [\left( \rtinv \right)_{(1)}
      \right ]^\prime (x_R)  + \left( \rtinv
    \right)_{(0)}(x_R)}_{{\cal O} (N_C^0)} + {\cal O} (N_C^{-1}) 
 =
  \underbrace{-\ui \rho(x_R)}_{{\cal O} (N_C^0)} + {\cal O}
  (N_C^{-1}) \, , 
\end{eqnarray}
At Leading Order (LO), we find
\begin{eqnarray}
\left( \rtinv \right)_{(1)}(x_R) = 0 \, . 
\end{eqnarray}
This forces $x_R$ to be real and guaranties that $m_R$ scales as
${\cal O} (N_C^0)$ in the $N_C\gg 3$ limit. At Next-to-Leading-Order
(NLO), we have
\begin{equation}
-\frac{{\rm Im}\ y_R}{N_C} = \rho(x_R) \left.\frac{1}{\frac{d}{ds}\left( \rtinv
 \right)_{(1)}   (s)}\right|_{s=x_R}  
\end{equation}
\begin{equation}
\frac{{\rm Re}\ y_R}{N_C} 
\left [\left( \rtinv \right)_{(1)} \right ]^\prime (x_R) =- \left(
\rtinv \right)_{(0)}(x_R) \, .
\end{equation}
Unitarity fixes the sign the of the imaginary part, showing that for
large, but finite $N_C$, the real pole comes from the 4th
quadrant. The first of the above equations ensures that the resonance
width, $\Gamma_R$, scales as ${\cal O} (N_C^{-1})$, for very large
values of $N_C$.

Now, we could rewrite Eq.~(\ref{eq:resonance}), with accuracy ${\cal
O} (N_C^{-2})$, as  
\begin{eqnarray} 
\rtinv (s_R) &=& \rtinv (m^2_R) - \ui m_R\Gamma_R \left[\rtinv
  \right]^\prime(m^2_R) -\frac{m_R^2\Gamma_R^2}{2}
\left[\rtinv \right]^{\prime\prime}(m^2_R) + {\cal O} (N_C^{-2})
\nonumber \\ &=& -\ui \rho(m_R^2) + m_R\Gamma_R
  \rho^{\,\prime}(m_R^2) + {\cal O} (N_C^{-2}) \, . \nonumber \\
\end{eqnarray}
 Thus, we find
that
\begin{eqnarray}
\rtinv (m^2_R) &=& \underbrace{\left.m_R\Gamma_R \left\{
  \rho^{\,\prime}+ \frac{m_R\Gamma_R}{2} \left[\rtinv
    \right]^{\prime\prime}\right\}\right|_{s=m^2_R}}_{{\cal O} (N_C^{-1})}
+ {\cal O} (N_C^{-3}) \, \\  m_R\Gamma_R &=&
\underbrace{\left.\frac{\rho}{\left[\rtinv \right]^{\prime}
  }\right|_{s=m^2_R}}_{{\cal O} (N_C^{-1})} + {\cal O} (N_C^{-3}) \, . 
\end{eqnarray}
Thus, at the resonance pole mass $\rtinv$ scales as ${\cal O}
(N_C^{-1})$ instead of ${\cal O} (N_C)$. The reason is that the pole
is moving, as we will show below, at speed $1/N_C^2$ towards the real
axis. This is the first theorem of this work. In principle, the
derivatives of $\rtinv$ at $s=m_R^2$ do still grow linearly with
$N_C$. On the other hand, since $\tan x = x+ {\cal O}(x^3)$, we also
find
\begin{eqnarray}
\delta(m^2_R) &=& \frac{\pi}{2} +
\underbrace{\delta^\prime(m^2_R)\left.\frac{\left[\rho^2\left[\rtinv
        \right]^{\prime}\right]^\prime}{2\left(\left[\rtinv
      \right]^{\prime}\right)^3}\right|_{s=m^2_R}}_{{\cal O}
  (N_C^{-1}) } + {\cal O} (N_C^{-3}) 
    \label{eq:pi2_R}
\end{eqnarray}
where we have used  that
\begin{eqnarray}
\delta^\prime(m^2_R) &=&
\left.\left[\frac{\rtinv}{\rho}\right]^\prime\frac{1}{1+
  \left(\frac{\rtinv}{\rho}\right)^2}\right|_{s=m^2_R}
=\underbrace{\left.\frac{\left[\rtinv \right]^{\prime}
  }{\rho}\right|_{s=m^2_R}}_{{\cal O} (N_C)} + {\cal O} (N_C^{-1}) \, . 
\end{eqnarray}
>From the above equations we see that $\delta^\prime(m^2_R)$ grows
linearly with $N_C$, while $\delta(m^2_R)$ reaches the value $\pi/2$,
up to corrections of the order ${\cal O} (N_C^{-1})$. This latter
result constitutes our second theorem. More importantly, from
Eq.~(\ref{eq:pi2_R}) it is trivial to find a value of $s$ for which
the phase shift differs of $\pi/2$ in terms suppressed by three powers
of the number of colors. This is to say
\begin{eqnarray}
\delta(s_{\rm BW}) = \frac{\pi}{2} +  {\cal O} (N_C^{-3}) \, ,  
\end{eqnarray}
where
\begin{eqnarray}
s_{\rm BW} & = & m^2_R -
\underbrace{\frac{\delta(m^2_R)-\pi/2}{\delta^\prime(m^2_R)}}_{{\cal
    O} (N_C^{-2}) } \, , \label{eq:bw} \\ & = &
m^2_R-\underbrace{\left.\frac{\left[\rho^2\left[\rtinv
        \right]^{\prime}\right]^\prime}{2\left(\left[\rtinv
      \right]^{\prime}\right)^3}\right|_{s=m^2_R}}_{{\cal O}
  (N_C^{-2})} + {\cal O} (N_C^{-4}) \, . 
\end{eqnarray}
Thus, we see that the existence of a pole in the SRS guaranties that
there exists a value of $s_{\rm BW}$, which can naturally be
identified with the BW position, where the phase shift is $\pi/2$, up
to ${\cal O} (N_C^{-3})$ corrections. The BW mass, $\sqrt{s_{\rm
    BW}}$, differs from the pole mass, $m_R$, in ${\cal O} (N_C^{-2})$
terms, which can be computed thanks to Eq.~(\ref{eq:bw}). Note that
the above relation has been deduced under the assumption of a finite
large $N_C$ limit of the resonance pole position. Our
Eq.~(\ref{eq:bw}) is nothing but the first iteration in Newton's
method for solving the BW condition, $\delta(s)=\pi/2$, starting from
the resonance mass as the initial guess. The meaning is just that
large $N_C$ implies that a straight line extrapolation of the phase
shift from the resonance should give a good estimate for the BW mass.


Ideally one would like to test Eq.~(\ref{eq:bw}) directly from data,
but the existing uncertainties in the resonance pole $m_R$, as well as
in the derivatives of the amplitude forced us to use instead a
suitable parameterization~\cite{Yndurain:2007qm,Caprini:2008fc}.  The
fact that the corrections are largely suppressed at large $N_C$
provides some confidence on the accuracy of the result. Using the
conformal mapping parameterizations of Ref.~\cite{Yndurain:2007qm} for
the isoscalar--scalar $\pi\pi$ phase shift\footnote{We use here the
ghost-full version and the Adler zero located at the lowest order ChPT
$s_A=m^2/2$~\cite{Yndurain:2007qm}.  Our results show little
dependence on this choice.}
\begin{eqnarray}
\rho (s) \cot \delta_{00}(s) = \frac{m^2}{s-m^2/2} 
\left[
  \frac{m}{\sqrt{s}}+ B_0 + B_1 w + B_2 w^2 \right] \, ,  \nonumber \\ 
\label{eq:conformal} 
\end{eqnarray} 
where the conformal mapping is 
\begin{eqnarray}
w(s)= \frac{\sqrt{s}-\sqrt{4m_K^2-s}}{\sqrt{s}+\sqrt{4m_K^2-s}} \, .
\end{eqnarray} 
We take three representative sets discussed in
Ref.~\cite{Yndurain:2007qm} and compiled for completeness in
Table~\ref{tab:table_conformal}.  We confirm that the resulting
complex pole position slightly overshoots the Roy equation value
$\sqrt{s_\sigma}= 441^{+16}_{-8} - \ui 272^{+9}_{-12} {\rm
  MeV}$~\cite{Caprini:2005zr}. Quoted errors in both the true BW
position $\delta_{00}(m_{\sigma,\ {\rm BW}}^2)=\pi/2$ and the
large-$N_C$ {\it predicted} BW position using Eq.~(\ref{eq:bw}) just
reflect uncertainties in the input parameters $B_0$, $B_1$ and $B_2$
as well as the induced complex pole. Actually, for Sets A and C we
find a $100\ {\rm MeV}$ wide stability plateau of the predicted value
from Eq.~(\ref{eq:bw}) around the pole mass.  The discrepancy is
compatible with the expected $1/N_C^4$ correction of the BW value,
given the fact that $\Gamma_\sigma$ is large. While a serious attempt
to evaluate this correction would require a much more reliable
parameterization and better data (higher order derivatives of the
phase shift enter) it is surprising that despite its large width, our
Eq.~(\ref{eq:bw}) may accommodate the large shift from the pole to the
BW mass by appealing large $N_C$ arguments which tacitly assume the
expansion of Eq.~(\ref{eq:sRII}) and thus a small width
approximation. Of course, the final answer regarding how the scalar
meson mass scales with $N_C$ can only be given by performing dynamical
lattice QCD calculations with variable $N_C$ (see
e.g. Ref. ~\cite{Teper:2008yi} for a review).

\begin{table}[ttt] 
\begin{center} 
\begin{tabular}{|c|c|c|c|}
\hline 
& Set A $\,$& Set B $\,$& Set C $\,$\\ 
\hline $B_0$ & 3.57(17) &  7.63(23) &  4.3(3)  \\ 
$B_1$ & $-$24.3(5) & $-$23.2(6) & $-$26.7(6) \\ 
$B_2$ & $-$6.3(1.3) & $-$23.0(1.4) & $-$14.1(1.4)  \\ 
$\sqrt{s_\sigma}\ [{\rm MeV}]$ 
    &  $\,$  466(4)$-$\ui 232 (3) $\,$ & $\,$ 477(7)$-$\ui 322(6) $\,$ & $\,$    476(6) $-$\ui 255(4) $\,$ \\ 
 $m_{\sigma,\ R}\ [{\rm MeV}]$   &    404(5)             &     350(10)             &   401(8)          \\
\hline 
$m_{\sigma,\ {\rm BW}}\ [{\rm MeV}]$  & 803 (4) & 865  (4) & 807  (5) \\ 
$m_\sigma|_{{\rm BW}, N_C\gg 3}\ [{\rm MeV}]$  & 657  (4) & 726  (5) &
678  (6) \\ 
\hline
\end{tabular}
\end{center} 
\caption{\label{tab:table_conformal} Large $N_C$ Predicted
  Breit-Wigner Resonances for the isoscalar-scalar channel
  $(T,J)=(0,0)$ in $\pi\pi$ scattering using the large $N_C$ formula
  $m_\sigma^2|_{\rm BW, large N_C} = m_{\sigma,R}^2 -
  (\delta_{00}(m_{\sigma,R}^2)-\pi/2)/\delta_{00}'(m_{\sigma,R}^2) $
  compared to the true BW result, $\delta_{00}(m_{\sigma, \ {\rm
  BW}}^2)=\pi/2$ where $s_\sigma = m_{\sigma,R}^2 - i m_{\sigma,R}
  \Gamma_{\sigma,R}$ represents the pole of the $S-$matrix in the SRS,
  $1/S_{\rm II}(s_\sigma)=0$. We use the parameterization for the
  $\delta_{00}(s)$ phase shift of Ref.~\cite{Yndurain:2007qm} (see
  Eq.~(\ref{eq:conformal}) in the main text).  }
\end{table}

Determining $s_R$ from $s_{\rm BW}$ or viceversa are in principle
equivalent procedures, but low energy based approximations such as
unitarized ChPT~\cite{Gasser:1983yg,Gasser:1984gg} are expected to
work better when predicting $s_{\rm BW}$ from $s_R$ since
$\sqrt{|s_R|} \sim 0.5\ {\rm GeV} $ and $ \sqrt{s_{\rm BW}} \sim 0.8
\ {\rm GeV}$. Actually, if we take the analytical one loop partial
wave amplitudes given in Ref.~\cite{Nieves:1999bx}, unitarize with the
IAM
method~\cite{Truong:1988zp,Dobado:1989qm,Dobado:1996ps,Hannah:1997ux}
and use $\bar l_1=-0.4(6)$, $\bar l_2=4.3(1)$, $\bar l_3=2.9(2.4)$,
$\bar l_4=4.4(2)$, from the analysis of Roy equations within
ChPT~\cite{Colangelo:2001df}\footnote{We have not considered here any
  type of statistical correlations. For a detailed discussion on
  effects due to them see Ref.~\cite{Nieves:1999zb}.}, we find a
reasonable good description of the phase shift at low energy that
leads to a rather good value of $\sqrt{s_\sigma} = 410(10)-\ui 270(10)
\ {\rm MeV}$. However, discrepancies with data become important as the
energy increases. and the phase shift never takes the value
$\delta_{00}(s)=\pi/2$. Despite these deficiencies, the large $N_C$
formula, Eq.~(\ref{eq:bw}), provides still a reasonable value for the
Breit-Wigner mass $m_\sigma|_{{\rm BW}, N_C\gg 3}= 600 (10) \ {\rm
  MeV}$.  The difference of this value to the estimate of
Table~\ref{tab:table_conformal} is consistent with the corresponding
values of the phase shifts since at $\sqrt{s}=500 \ {\rm MeV}$ one has
$\delta_{00} =45.7(6)^{\rm o}, 39.1(6)^0$ and $ 43.4(9)^{\rm o}$ for
Sets A,B and C respectively whereas one finds a significant smaller
value $\delta_{00}= 34.7(5)^{\rm o}$ for the chiral IAM unitarized
case with $\bar l_{1,2,3,4}$ from Ref.~\cite{Colangelo:2001df}. We
stress that the phase shift in the chiral unitary representation
itself never passes through $90^0$.  This result reinforces the
advocated picture; while in terms of the chiral representation the
pole and the BW masses are far apart, within the large $N_C$ framework
they are connected, as they approach to each other at speed ${\cal
  O}(1/N_C^2)$. Note, however, that in practice we {\it never} depart
from the physical $N_C=3$ value.

We summarize our results. We have analyzed the connection between the
pole mass and the Breit-Wigner mass of the $\pi\pi$ scattering
amplitude within the large $N_C$ expansion. We have shown that
assuming that both masses are ${\cal O}(N_C^0)$ the difference is
${\cal O}(N_C^{-2})$ {\it parametrically} suppressed and computable
numerically from the data. This allows to {\it predict} the BW mass
from the pole mass successfully even in the hostile case of the rather
wide $f_0(600)$ resonance. Thus, while the pole and Breit-Wigner
masses are far apart numerically they turn out to be connected within
the large $N_C$ approximation. That would indicate the presence of
a $q\bar q$ component in the $\sigma-$wave function. Such component,
likely sub-dominant in the real world $N_C=3$~\cite{Pelaez:2006nj},
would ensure for a sufficiently large number of colours, the
  $N_C-$behaviour ($m_\sigma \sim N_C^0$ and $\Gamma_\sigma \sim
1/N_C$) of the $\sigma$ pole parameters that has allowed us to relate
pole and BW masses.



\end{document}